\documentclass{article}

\usepackage{arxiv}

\usepackage[utf8]{inputenc} 
\usepackage[T1]{fontenc}    
\usepackage{hyperref}       
\usepackage{url}            
\usepackage{booktabs}       
\usepackage{amsfonts}       
\usepackage{nicefrac}       
\usepackage{microtype}      
\usepackage{graphicx}
\usepackage[round]{natbib}  
\providecommand{\tightlist}{\setlength{\itemsep}{0pt}\setlength{\parskip}{0pt}} 
\graphicspath{ {./} }
\renewcommand{\today}{} 
\title{metasignal: A Python Package for Comprehensive Metacognitive Analysis and Decision-Making\thanks{This is a preprint. The corresponding software paper has been submitted to the Journal of Open Source Software (JOSS).}}

\author{
 Saurabh Ranjan \\
  University of Florida\\
  Gainesville, FL, USA \\
  \texttt{ranjan.saurabh@outlook.com} \\
  \And
 Mukesh Makwana \\
  Brown University\\
  Providence, RI, USA \\
  \texttt{mukesh\_makwana@brown.edu} \\
  \And
 Konstantina Sokratous \\
  University of Missouri\\
  Columbia, MO, USA \\
  \texttt{ksokratous@missouri.edu} \\
  \And
 Brian Odegaard \\
  University of Florida\\
  Gainesville, FL, USA \\
  \texttt{bodegaard@ufl.edu} \\
}

\begin{document}
\maketitle

\begin{abstract}
\texttt{metasignal} is an open-source Python package for signal detection theory (SDT) and metacognitive measurement. It implements the 17 metacognitive measures evaluated by \citet{rahnev2025}, together with the reference variables \emph{d'} (perceptual sensitivity), response criterion \emph{c} (response bias), and mean confidence. The 17 measures comprise three meta-d'-family estimates, \emph{meta-d'}, \emph{M-ratio}, and \emph{M-difference}; four nonparametric Type-2 measures, the Type-2 area under the receiver-operating-characteristic curve (\emph{AUC2}), \emph{Gamma}, \emph{Phi}, and \emph{delta confidence}, together with their eight SDT-normalized ratio and difference forms; and two model-based measures, \emph{meta-noise} and \emph{meta-uncertainty}. A single function computes the complete set from trial-level stimulus, response, and confidence arrays. \texttt{metasignal} currently supports binary (two-alternative) discrimination tasks, in which each trial's stimulus and response are coded with exactly two categories. The package also provides a command-line interface, group summaries, bootstrap confidence intervals, permutation tests, optional hierarchical Bayesian models, and information-theoretic measures. \texttt{metasignal} unifies these measures in a single platform to encourage broader metacognition research and adoption in decision-making studies.
\end{abstract}


\section{Statement of Need}\label{statement-of-need}

Metacognition is the ability to evaluate the quality of one's own decisions. It supports learning, adaptive choice, and communication of uncertainty \citep{flemingdolan2012, yeungsummerfield2012}. Metacognitive measurement is also relevant to clinical research \citep{hoven2019}, behavioral economics \citep{moorehealy2008}, and the evaluation of uncertainty reported by artificial-intelligence systems, where metacognitive sensitivity has been proposed as key to calibrating human trust in AI \citep{steyvers2025, griot2025, lee2025trust}. A consensus statement identified reliable computational models and standardized measurement as major goals for visual-metacognition research \citep{rahnev2022}.

Numerous metacognitive measures have been proposed, but they quantify distinct properties and vary in their sensitivity to task performance, response bias, confidence bias, and sample size. Among these, meta-d' is widely used because it expresses metacognitive sensitivity in signal detection theory (SDT) units directly comparable to perceptual sensitivity \citep{maniscalco2012}. In plain terms, it measures how well confidence tracks actual accuracy, on the same scale researchers already use for perceptual sensitivity. Although the original MATLAB implementation \citep{maniscalco2014} and its Python port \citep{lee_metad_python} brought meta-d' to a broad audience, both tools focus exclusively on this single measure. The other sixteen measures benchmarked by \citet{rahnev2025} remained scattered across isolated codebases, leaving no single Python package that spans the full set. This fragmentation has hindered researchers from comparing metacognitive measures within a single, reproducible workflow.

\citet{rahnev2025} provided the first broad empirical comparison of 17 measures. The study assessed validity and precision, dependence on nuisance variables, split-half reliability, and test--retest reliability. \texttt{metasignal} translates this benchmark into a documented Python package that requires no proprietary software. It is intended for cognitive scientists, psychophysicists, neuroscientists, clinical researchers, and researchers studying confidence in artificial systems.

\section{State of the Field}\label{state-of-the-field}

The MATLAB toolbox by \citet{maniscalco2014} and its Python port \citep{lee_metad_python} remain important reference implementations for standard meta-d' estimation. Packages like \texttt{metadpy} \citep{legrand2021} and HMeta-d \citep{fleming2017} introduced valuable Bayesian and hierarchical estimation frameworks, though they remain focused primarily on the meta-d' family. In R, \texttt{statConfR} \citep{rausch2025} provides confidence models alongside information-theoretic metrics. While each of these tools excels within its intended scope, none offers the full battery of metrics benchmarked by \citet{rahnev2025} through a single, native Python interface.

\texttt{metasignal} addresses this gap by bringing an expansive collection of metacognitive measures under a single Python package umbrella. Rather than replacing existing tools, it complements them by providing broad metric coverage and a standardized trial-level API, which together lower the barrier to wider adoption and more rigorous metacognition research. The optional \texttt{sdtbayes} subpackage enables Stan-based hierarchical estimation, and the experimental \texttt{itmc} subpackage implements measures grounded in metacognitive information theory \citep{dayan2023, meyen2025}, with its \texttt{statconfr} backend ported from the R \texttt{statConfR} package \citep{rausch2025}.

\section{Software Design}\label{software-design}

\texttt{metasignal} is built around four primary design principles. First, input data (stimulus identity, behavioral response, and confidence) are structured as tabular trial-level data compatible with standard Python processing workflows. Second, the core \texttt{stdpy} layer relies solely on NumPy \citep{harris2020} and SciPy \citep{virtanen2020}, providing a pure-Python implementation. Third, it offers flexible execution: \texttt{compute\_all\_measures} provides a single entry point for broad profiling, while individual functions remain exposed for focused queries. Fourth, complex Bayesian dependencies are isolated from the lightweight frequentist core.

When invoked, \texttt{compute\_all\_measures} returns 26 values. The first 20 span d', decision criterion, mean confidence, and the 17 metacognitive metrics. The final six outputs (log-likelihood, AIC, BIC, AICc, fitted parameter count, and sample size) serve as diagnostic parameters for the meta-d' fit rather than distinct metacognitive measures. An optional \texttt{return\_type} argument (\texttt{'dict'} or \texttt{'dataframe'}) labels these same 26 values by measure name instead of position; the default \texttt{'array'} preserves positional, backward-compatible output.

\begin{figure}
\centering
\includegraphics[width=0.95\linewidth,height=\textheight,keepaspectratio,alt={Architecture of metasignal. Trial-level data enter the stable stdpy layer. Analysis and command-line layers provide inference and batch use; Bayesian and information-theoretic components are optional.}]{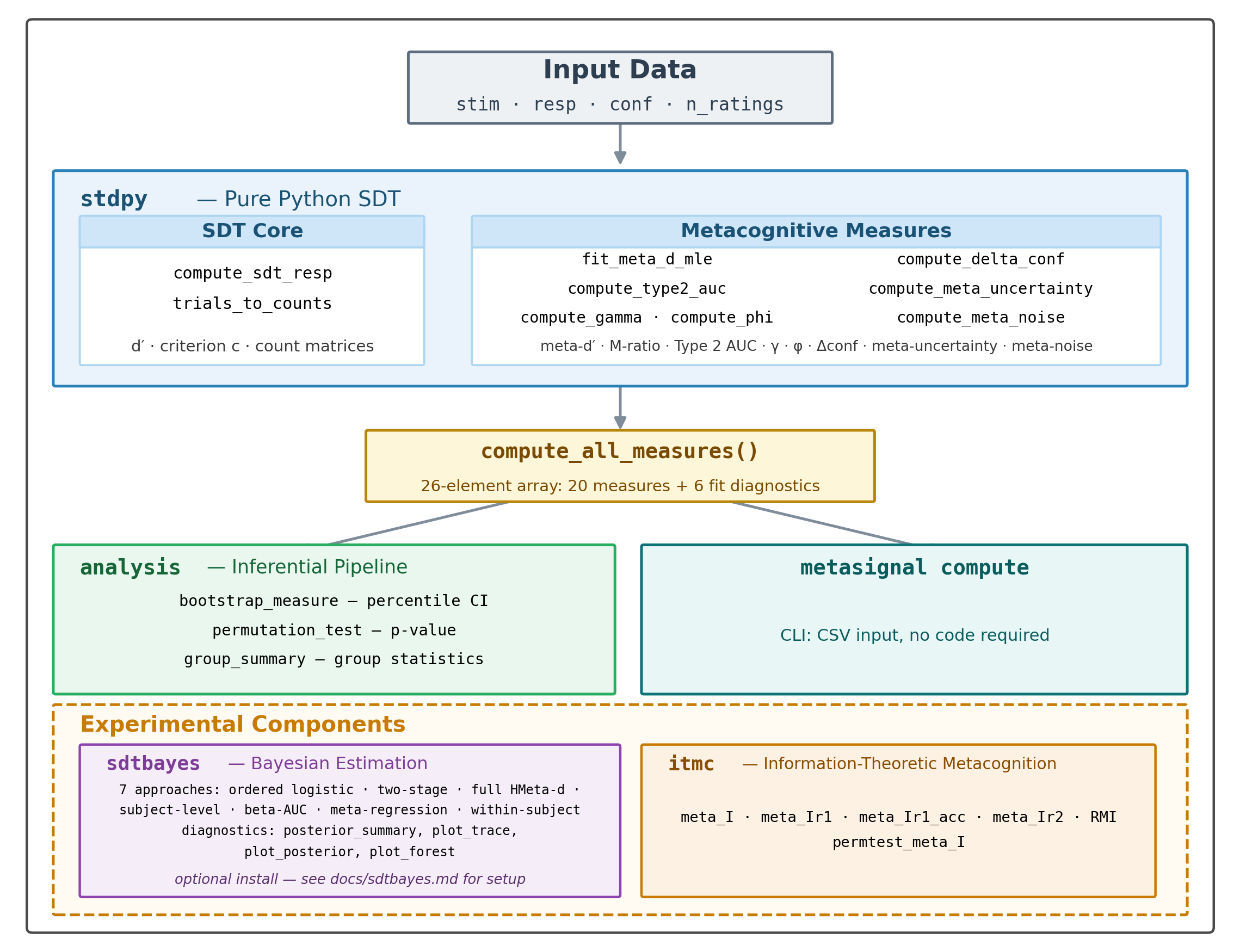}
\caption{\textbf{Architecture of \protect\texttt{metasignal}.} Trial-level stimulus, response, confidence, and rating-scale arrays (\texttt{stim}, \texttt{resp}, \texttt{conf}, \texttt{n\_ratings}) enter the stable \texttt{stdpy} core, which computes Type-1 SDT quantities ($d'$, criterion $c$) and all 17 metacognitive measures; \texttt{compute\_all\_measures} packages these into a single 26-element array (20 measures plus 6 meta-$d'$ fit diagnostics). This output feeds the \texttt{analysis} layer (bootstrap confidence intervals, permutation tests, group summaries) and the \texttt{metasignal compute} command-line interface, which accepts CSV input without requiring code. The dashed box shows optional, dependency-isolated extensions: \texttt{sdtbayes} for hierarchical Bayesian estimation (seven modeling approaches) and \texttt{itmc} for information-theoretic metacognition measures.}
\end{figure}

\section{Methods}\label{methods}

\subsection{Core computation}\label{core-computation}

The package first converts trial-level data into Type-2 response-count arrays. Here, ``Type-1'' denotes the original stimulus/response decision and ``Type-2'' the confidence judgment about it, unrelated to Type I/Type II statistical error. Type-1 sensitivity and criterion are computed under the equal-variance SDT model \citep{green1966}. Nonparametric Type-2 measures are computed directly from the observed counts. Meta-d' is estimated by maximum likelihood \citep{maniscalco2012}, from which M-ratio and M-difference are derived. SDT-normalized Ratio and Difference measures compare observed statistics with statistics expected under an ideal equal-variance SDT observer.

Meta-noise is estimated with the lognormal confidence-noise model used by the MATLAB benchmark. The Python implementation follows the MATLAB procedure: it evaluates the zero-noise Gaussian baseline, searches outward from that lower bound, uses golden-section optimization, and evaluates the precomputed integral table using inverse-distance weighting. Meta-uncertainty is fitted with its corresponding model-based estimator.

\subsection{Validation data and procedure}\label{validation-data-and-procedure}

Validation used the six datasets distributed with the Rahnev analysis pipeline: Haddara, Maniscalco, Rouault experiments 1 and 2, Shekhar, and Locke. The comparison used three sources:

\begin{enumerate}
\def\labelenumi{\arabic{enumi}.}
\tightlist
\item
  values reported in the text, figures, and supplementary tables of \citet{rahnev2025};
\item
  subject-level MATLAB arrays stored in \texttt{matlab/metasignal\_mat/Results}; and
\item
  Python arrays generated from the same trial subsets.
\end{enumerate}

Ten subject-level analysis arrays were compared: raw estimates, metacognitive-bias recoding, odd--even splits, difficulty conditions, and response-bias conditions. For each measure we examined Pearson correlation, mean absolute error, root-mean-square error, signed bias, maximum absolute error, and missing-value agreement. We also reproduced 19 statistical checks from the supplementary analyses and compared the 17-measure profiles for task-performance dependence, metacognitive-bias dependence, response-bias dependence, and test--retest reliability. The test--retest comparison used the same bin size (400 trials), day structure, and Fisher-\emph{z} aggregation as the MATLAB scripts. Precision was also recomputed directly from raw trial data using the bin-stratified protocol \citet{rahnev2025} describes (non-overlapping 50-, 100-, 200-, and 400-trial bins), under a bin-instance cap required for computational tractability (see Limitations).

The complete workflow is implemented in \texttt{analysis/rahnev\_comparison/scripts}. The JSON results, CSV summaries, identity plots, profile overlays, and combined PDFs are stored beside the scripts. This makes the validation repeatable rather than dependent on a manually prepared table.

\section{Validation Results}\label{validation-results}

\begin{figure}
\centering
\includegraphics[width=0.88\linewidth,height=\textheight,keepaspectratio,alt={Comparison of published Rahnev values with MATLAB and Python replications. Panels show task-performance effects, metacognitive-bias effects, response-bias correlations, test--retest ICC, and the task-performance profile across all 17 measures.}]{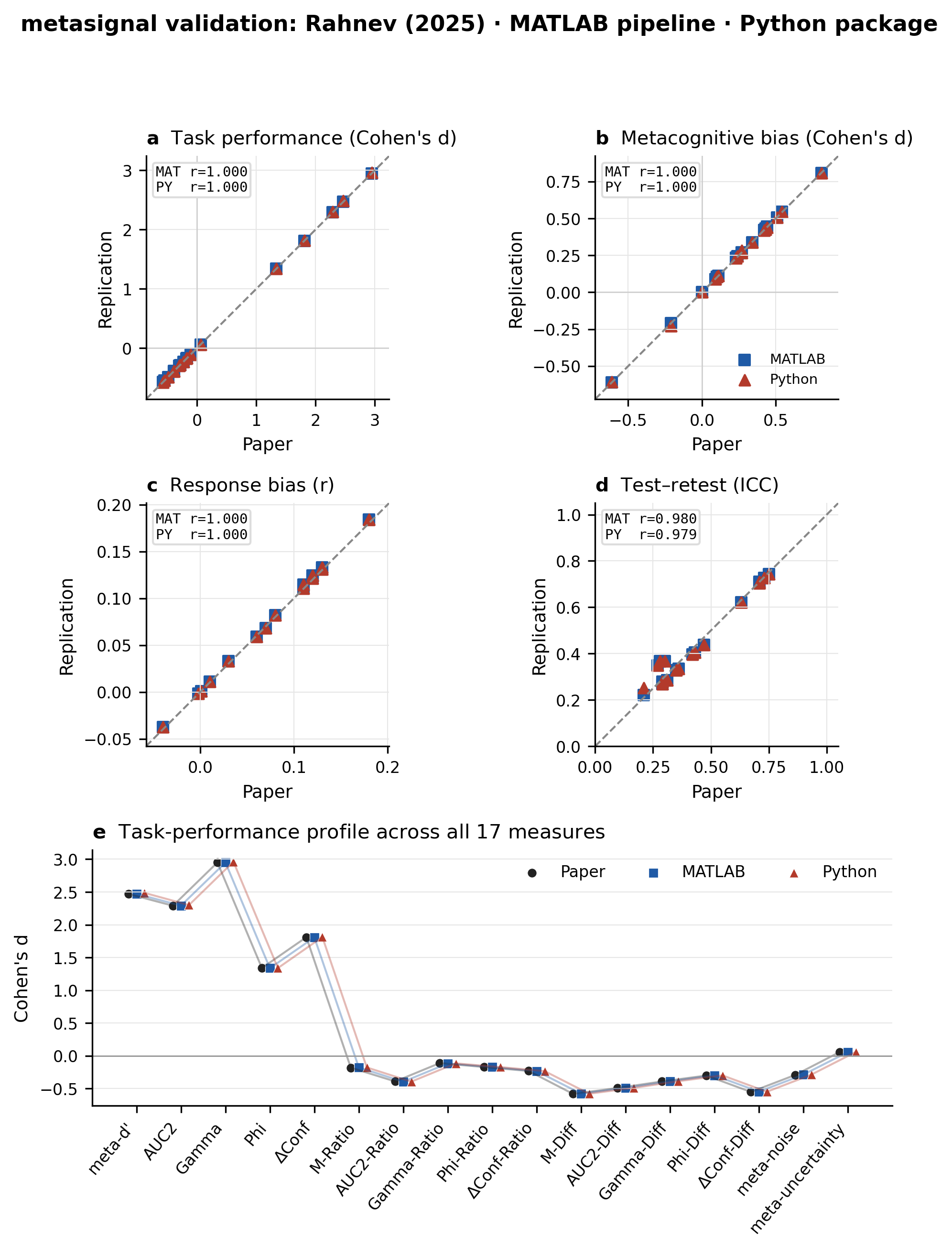}
\caption{\textbf{Cross-implementation validation against \citet{rahnev2025}.} In each panel, published values (x-axis, ``Paper'') are plotted against replication values (y-axis) from the MATLAB pipeline (blue squares) and the Python \texttt{metasignal} package (red triangles); the dashed line marks exact identity. Published values in all five panels are the per-measure averages reported in \citet{rahnev2025}'s Figure 7 (``Summary of results''). MAT and PY report the Pearson correlation between each replication and the published values. \textbf{a}, Task-performance effect sizes (Cohen's \emph{d}; MAT \emph{r} = 1.000, PY \emph{r} = 1.000). \textbf{b}, Metacognitive-bias effect sizes (Cohen's \emph{d}; MAT \emph{r} = 1.000, PY \emph{r} = 1.000). \textbf{c}, Response-bias correlations (\emph{r}; MAT \emph{r} = 1.000, PY \emph{r} = 1.000). \textbf{d}, Test--retest reliability (intraclass correlation coefficient, ICC; MAT \emph{r} = 0.980, PY \emph{r} = 0.979). \textbf{e}, Task-performance profile (Cohen's \emph{d}) across all 17 metacognitive measures, comparing published values (black circles) with MATLAB (blue squares) and Python (red triangles) replications.}
\end{figure}

All 10 subject-level comparison arrays passed the predefined comparison gate, and all 19 supplementary statistical checks matched in statistical significance and reference \emph{t} value. These 19 checks are the paired-contrast \emph{t}-values reported in \citet{rahnev2025}'s Supplementary Tables 3, 4, and 6 (Shekhar difficulty, Rouault1 difficulty, and Haddara metacognitive-bias contrasts, respectively; 8 + 4 + 7 = 19 measure-level values), reproduced independently in both the MATLAB and Python pipelines — distinct from the profile-level correlations reported in the paragraphs below. Full results are in \texttt{analysis/rahnev\_comparison/ANALYSIS\_REPORT.md} in the repository. Across the 18 non-model-based measures, MATLAB and Python agreed to numerical precision at the subject level (maximum systematic bias below $1.5 \times 10^{-3}$; per-analysis correlations of 1.000). Ten of these measures agreed to within a maximum absolute difference of $10^{-2}$ across every subject and analysis.

The analysis profiles were essentially identical across sources: task-performance dependence (paper--Python \emph{r} = 1.000; MATLAB--Python \emph{r} = 1.000), metacognitive-bias dependence (paper--Python \emph{r} = 1.000; MATLAB--Python \emph{r} = 1.000), and response-bias dependence (paper--Python \emph{r} = 1.000; MATLAB--Python \emph{r} = 1.000). Test--retest ICC (intraclass correlation coefficient), matched to the MATLAB bin-size-400 protocol, also aligned closely (MATLAB--Python \emph{r} = 0.999).

Some differences remain and are reported as limitations rather than hidden. Meta-d', M-ratio, and M-difference correlate at 1.000 across analyses but can differ by up to about 0.14 for individual participants because both pipelines use bounded numerical maximum-likelihood optimizers with different internal solvers. Meta-noise and meta-uncertainty retain a small number of subject-level differences in the sparsest difficulty subsets, where the likelihood surface is flat. A single degenerate value returned by the MATLAB meta-noise search (near 0.4959) is identified programmatically and excluded from agreement statistics. For users who need to reproduce a specific MATLAB result file rather than a scientifically stable estimate, the model-based estimators expose an optional \texttt{matlab\_compat} mode that mimics MATLAB's single-start optimizer configuration. The package therefore supports the scientific conclusions of \citet{rahnev2025} and reproduces the MATLAB pipeline to numerical precision for the non-model-based measures, while documenting the residual, well-understood variation in low-information model fits.

The experimental \texttt{itmc} subpackage was validated separately from the core \texttt{stdpy} benchmark above. Its \texttt{statconfr} backend reproduces the R \texttt{statConfR} package's \texttt{estimateMetaI()} output to machine precision on synthetic data spanning a range of sensitivity, bias, and confidence-scale configurations ($r = 1.000$ for all five measures under deterministic computation), and independently reproduces a hand-worked numerical example from \citet{dayan2023} exactly (meta-I = 0.094). Full results, including a known ratio-measure instability at low sensitivity common to both implementations, are in \texttt{analysis/itmc\_comparison/}.

\section{Simple Procedure for Use}\label{simple-procedure-for-use}

\subsection{Installation}\label{installation}

Install from the public repository:

\begin{verbatim}
pip install git+https://github.com/saurabhr/metasignal.git
\end{verbatim}

For development or reproducible validation:

\begin{verbatim}
git clone https://github.com/saurabhr/metasignal.git
cd metasignal
pip install -e .
\end{verbatim}

\subsection{Prepare trial-level data}\label{prepare-trial-level-data}

Each trial requires four values:

\begin{itemize}
\tightlist
\item
  \texttt{stim}: stimulus category, coded with two values such as 0 and 1;
\item
  \texttt{resp}: participant response, coded with the same two values;
\item
  \texttt{conf}: integer confidence rating from 1 to \texttt{n\_ratings}; and
\item
  \texttt{n\_ratings}: number of possible confidence levels.
\end{itemize}

The arrays must have the same length. Missing trials are removed jointly. A study should compute participant-level estimates separately before group inference.

\subsection{Compute all measures}\label{compute-all-measures}

The following self-contained example simulates one participant with a fixed random seed, so that any user can run it and obtain the same numbers. This doubles as a minimal reproducibility check of the installation.

\begin{verbatim}
import numpy as np
from metasignal import stdpy

rng = np.random.default_rng(2025)
n = 800
stim = rng.integers(0, 2, n)                 # two stimulus categories
evidence = (stim * 2 - 1) * 0.9 + rng.normal(0, 1, n)
resp = (evidence > 0).astype(int)            # observer response
edges = np.quantile(np.abs(evidence), [0.25, 0.5, 0.75])
conf = np.digitize(np.abs(evidence), edges) + 1   # 4-point confidence

values = stdpy.compute_all_measures(stim, resp, conf, n_ratings=4)

meta_d  = values[0]    # meta-d'
auc2    = values[1]    # Type-2 AUC
m_ratio = values[5]    # M-ratio (meta-d'/d')
dprime  = values[17]   # d'
print(f"d'={dprime:.3f}  meta-d'={meta_d:.3f}  M-ratio={m_ratio:.3f}  AUC2={auc2:.3f}")
\end{verbatim}

Running this prints:

\begin{verbatim}
d'=1.929  meta-d'=1.841  M-ratio=0.955  AUC2=0.762
\end{verbatim}

\texttt{compute\_all\_measures} returns all 26 values in a fixed order; the first 20 are the 17 metacognitive measures plus d', criterion, and mean confidence, and the last six are meta-d' fit diagnostics. Researchers should use realistic trial counts for scientific estimation; the seed here only makes the example reproducible.

\subsection{Compute selected measures}\label{compute-selected-measures}

\begin{verbatim}
dprime, criterion, ln_beta = stdpy.compute_sdt_resp(stim, resp)
nr_s1, nr_s2 = stdpy.trials_to_counts(stim, resp, conf, n_ratings=4)
meta_fit = stdpy.fit_meta_d_mle(nr_s1, nr_s2)
auc2 = stdpy.compute_type2_auc(nr_s1, nr_s2)

print(meta_fit["meta_da"])
print(meta_fit["M_ratio"])
\end{verbatim}

\subsection{Command-line use}\label{command-line-use}

\begin{verbatim}
metasignal compute \
  --stim "0,1,0,1,0,1,0,1" \
  --resp "0,1,0,0,0,1,1,1" \
  --conf "4,4,3,1,3,4,2,3" \
  --n-ratings 4
\end{verbatim}

Trial data can also be read from a CSV with one trial per row instead of typed inline:

\begin{verbatim}
cat > trials.csv <<'CSV'
stim,resp,conf
0,0,4
1,1,4
0,0,3
1,0,1
0,0,3
1,1,4
0,1,2
1,1,3
CSV

metasignal compute --csv trials.csv --n-ratings 4
\end{verbatim}

Both commands above print the identical 26-value table (they encode the same eight trials); this was verified directly rather than assumed. Column names default to \texttt{stim}, \texttt{resp}, and \texttt{conf} and can be overridden with \texttt{-\/-stim-col}, \texttt{-\/-resp-col}, and \texttt{-\/-conf-col} for CSVs with different headers.

\subsection{Recommended scientific workflow}\label{recommended-scientific-workflow}

\begin{enumerate}
\def\labelenumi{\arabic{enumi}.}
\tightlist
\item
  Check that stimulus and response each contain two categories.
\item
  Confirm that confidence values are integers within the declared scale.
\item
  Compute measures independently for each participant and condition.
\item
  Inspect missing values and convergence diagnostics before group analysis.
\item
  Use bootstrap intervals or permutation tests when sampling distributions are uncertain.
\item
  Report the measure definition, trial count, confidence scale, exclusion criteria, and reliability protocol.
\item
  For direct replication, preserve the original bin sizes, recoding rules, and random-resampling procedure.
\end{enumerate}

\section{Research Impact Statement}\label{research-impact-statement}

\texttt{metasignal} turns a broad methodological benchmark into reusable research infrastructure. It provides a common implementation for comparing measures, lowers the barrier for laboratories without MATLAB, and makes numerical validation visible and reproducible. Seven tutorial notebooks cover preprocessing, measure computation, statistical tables, difficulty dependence, bias, split-half reliability, and test--retest reliability. Automated tests, continuous integration, API documentation, and contribution guidelines support reuse and extension.

The validation materials are also a scholarly contribution. They identify implementation details that materially affect results, especially the zero-noise boundary in meta-noise fitting and the use of proportions in SDT-expected ratings. Recording these details helps prevent apparently contradictory results that arise from software rather than theory.

\section{Limitations}\label{limitations}

\subsection{Validation Scope}\label{validation-scope}

No single metacognitive measure is optimal for every experimental design \citep{rahnev2025}. Model-based and ratio measures can become unstable with small sample sizes or sparse confidence categories, particularly when expected denominators approach zero. Users should select measures based on their scientific questions and report analysis settings completely.

While \texttt{metasignal} perfectly reproduces the core task-performance, metacognitive-bias, and response-bias benchmark profiles from \citet{rahnev2025} ($r = 1.000$), precision validation requires nuanced interpretation. Precision was recomputed directly from raw trial data using the bin-stratified protocol \citet{rahnev2025} describes (non-overlapping 50-, 100-, 200-, and 400-trial bins). Under a bin-instance cap required for computational tractability, precision estimates display elevated sampling noise, occasionally causing unstable measures (e.g., Gamma-Ratio) to invert signs. Consequently, current precision figures should be treated as illustrative rather than an exact protocol match to \citet{rahnev2025}.

\section{Conclusions}\label{conclusions}

\texttt{metasignal} provides a simple, open, and validated Python interface to the principal SDT and metacognitive measures compared by \citet{rahnev2025}. That numerical fidelity means researchers can adopt \texttt{metasignal} in place of the MATLAB pipeline without sacrificing comparability to the published benchmark. The package combines breadth, ease of use, transparent validation, and optional inferential tools, and remaining differences are localized to low-information model fits or unmatched reliability protocols and are documented explicitly. Taken together, these properties make \texttt{metasignal} suitable for reproducible metacognition research while preserving appropriate scientific caution.

\section*{Code Availability}

\texttt{metasignal} is open source (MIT license) and available at \url{https://github.com/saurabhr/metasignal}. Full documentation, including the API reference and tutorials, is hosted at \url{https://metasignal.readthedocs.io/en/latest/}. The complete \citet{rahnev2025} replication and validation workflow (comparison scripts, JSON/CSV results, and figures) is included in the repository under \texttt{analysis/rahnev\_comparison/}, and the bundled MATLAB reference pipeline and results are under \texttt{matlab/metasignal\_mat/}. The original MATLAB code for the 17 metacognition measures, from which the bundled MATLAB pipeline is derived, is available at \url{https://osf.io/y5w2d/}\citep{rahnev2025}.

\section*{AI Usage Disclosure}

Claude (Anthropic) and OpenAI language models were used to assist with initial drafts of portions of the source code, documentation, validation report, and manuscript. All software behavior, numerical comparisons, citations, scientific claims, and final wording were reviewed and edited by the authors. Responsibility for the software and manuscript remains with the authors.

\bibliographystyle{unsrtnat}
\bibliography{references}

\end{document}